\documentstyle[11pt]{article}
\input epsf

\newcommand\lsim{\mathrel{\rlap{\lower4pt\hbox{\hskip1pt$\sim$}}
    \raise1pt\hbox{$<$}}}
\newcommand\gsim{\mathrel{\rlap{\lower4pt\hbox{\hskip1pt$\sim$}}
    \raise1pt\hbox{$>$}}}

\begin{document}

\begin{flushright}
SUSSEX-AST 96/10-3\\
astro-ph/9610215\\
October 1996
\end{flushright}

\vspace*{24pt}

\begin{center}
{\Large INFLATION, STRUCTURE FORMATION AND DARK MATTER\footnote{To appear, 
proceedings of `Aspects of Dark Matter in Astro- and Particle Physics', 
Heidelberg, September 1996 (World Scientific).}}

\vspace*{24pt}

{\large ANDREW R.~LIDDLE AND PEDRO T.~P.~VIANA}

\vspace*{24pt}
{\em 
Astronomy Centre, University of Sussex\\ Brighton BN1 9QH~~~Great 
Britain}

\vspace*{24pt}
{\bf Abstract}\\
\end{center}

\begin{quote}
The formation of structure in the Universe offers some of the 
most powerful evidence in favour of the existence of dark matter in the 
Universe. We summarize recent work by ourselves and our collaborators, using 
linear and quasi-linear theory to probe the allowed parameter space of 
structure formation models with perturbations based on the inflationary 
cosmology. Observations used include large and intermediate angle microwave 
background anisotropies, galaxy clustering, the abundance of galaxy clusters 
and object abundances at high redshift. The cosmologies studied include 
critical density models with cold dark matter and with mixed dark matter, 
cold dark matter models with a cosmological constant and open cold dark 
matter models. Where possible, we have updated results from our journal 
papers.
\end{quote}

\section{Introduction}

The demise of the standard cold dark matter (CDM) model, and the wide range 
of minor variants which have surfaced in its wake, have forced us to come to 
terms with the fact that structure formation models depend on a wide range 
of parameters. Cosmological parameters, such the Hubble parameter, the 
density parameter and a possible cosmological constant, all have a 
significant effect on the evolution of structure. So too does the assumed 
matter content; almost all models contain cold dark matter, but we must also 
worry about the appropriate value of the baryon density, the possibility 
that some of the dark matter may be hot, and even the present energy density 
in the form of massless species. To add to that, there is uncertainty 
concerning the form of the initial perturbations from which structure grows; 
for example, even the simplest versions of the inflationary universe 
paradigm typically generate more complicated spectra than the 
scale-invariant Harrison--Zel'dovich spectrum, giving yet further freedom, 
whilst topological defect models such as cosmic strings lead to a yet more 
complex situation.

Such a wide parameter space is not amenable to detailed investigation via 
numerical simulations, which provide the most rigorous test of cosmological 
models. In order to discover the parameters best suited to matching the 
observational data, and as importantly to determine which possibilities are 
definitely excluded, it is necessary to use more economical techniques. In 
that vein, we and our collaborators have recently been investigating 
the parameter space through the use of linear and quasi-linear perturbation 
theory, which allows a semi-analytical comparison of theory against a range 
of observations. We have concentrated on constraints which can be applied 
directly to the matter and radiation power spectra, rather than becoming 
involved in the delicate issues of biased galaxy formation introduced when 
the galaxy correlation function is studied in detail. This has enabled us to 
carry out an investigation of wide areas of parameter space, which we have 
published in a series of papers. These cover the case of a critical density 
universe~\cite{LLSSV} (both the cold dark matter and mixed dark matter 
cases), a flat universe with a cosmological constant~\cite{LLVW} and an open 
universe \cite{LLRV}. Finally, we returned to the case of critical density 
cold dark matter models~\cite{WVLS} to investigate whether their perilous 
position could be alleviated by allowing a substantial increase to the 
baryon density. Throughout this work, we have placed particular focus on 
taking the inflationary paradigm seriously, and consequently considering a 
range of possible initial power spectra.

In this brief article there is no room to give the full details of these 
papers. Instead, we aim to give an overview of some of the 
possibilities. Where possible we have taken the opportunity to update 
constraints from our published work, especially with regard to intermediate 
scale microwave background anisotropies. The present situation is described 
in great detail in Viana's thesis \cite{thesis}. An extensive list of 
references can be found in the papers cited here.

\section{Theoretical Input}

\subsection{Inflationary perturbations}

During an inflationary epoch, quantum fluctuations swept up in the rapid 
expansion lead to perturbations in the present universe. A review 
of this mechanism, and of the modelling of an inflationary epoch, is 
given in Ref.~6. Inflation produces not only a spectrum of 
density perturbations, but also one of gravitational waves. These latter are 
capable of influencing large-angle microwave background anisotropies, as 
witnessed by the {\it COBE} satellite, and must be taken into account in a 
full comparison with observations.

In absolute generality, inflationary models can lead to a wide range of 
different spectra of density perturbations. However, almost all models fall 
within the slow-roll paradigm, within which physical conditions only vary 
slowly as inflation proceeds. In that situation, one finds that the density 
perturbations can be accurately described by a power-law spectrum \cite{LL}, 
sometimes known as a tilted spectrum.\footnote{Often the tilt can be given 
the following interpretation. If inflation is to end, it must move away from 
the `scale invariance' of exponential expansion. Although perturbations on 
observable scales were generated some time before inflation came to an end, 
typically one does see the beginnings of the approach to the end of 
inflation, with the tilt arising in the incipient breaking of the expansion 
away from exponential.} We shall work within that regime.

Slow-roll inflation can therefore be characterized by three parameters. The 
overall amplitude of density perturbations has always been considered a free 
parameter in large-scale structure studies. Along with that comes the 
spectral index of the density perturbations $n$, and a parameter $R$ which 
measures the extent to which gravitational waves contribute to {\it COBE}. 
Since gravitational waves only affect the large-angle microwave background 
observations, only a single parameter need be introduced to quantify their 
effect.

For any given inflation model, $n$ and $R$ are easy to compute. The range of 
possible values is wide; indeed, wider than present observations permit. 
Consequently, some inflation models have already been excluded (extended 
inflation being the most notable example~\cite{LL}), a promising development 
we can expect to see much more of in the future. Given this, it is 
satisfactory to treat $n$ and $R$ as completely free parameters in seeking 
viable large-scale structure models. Needless to say, the extra freedom 
which they represent has a serious influence on what can be said concerning 
the other cosmological parameters and the nature of the dark matter.

\subsection{Cosmological parameters}

In our studies, we permitted variation of five cosmological parameters, in 
addition to the inflationary parameters. They are the Hubble parameter $h$, 
the density parameter $\Omega$, the cosmological constant $\Lambda$, the 
amount of hot dark matter $\Omega_{\nu}$ and the amount of baryonic 
matter $\Omega_{{\rm B}}$. We didn't attempt to consider all possible 
combinations however; we only considered a hot component in universes with 
critical density (since other models can comfortably fit the data without a 
hot component, and to have too many different components with similar 
densities seems contrived), and a cosmological constant was only introduced 
with the appropriate value to make the universe spatially flat. In all 
cases, cold dark matter was used to make up the required total density.

\begin{figure}[t]
\centering
\leavevmode\epsfysize=9cm \epsfbox{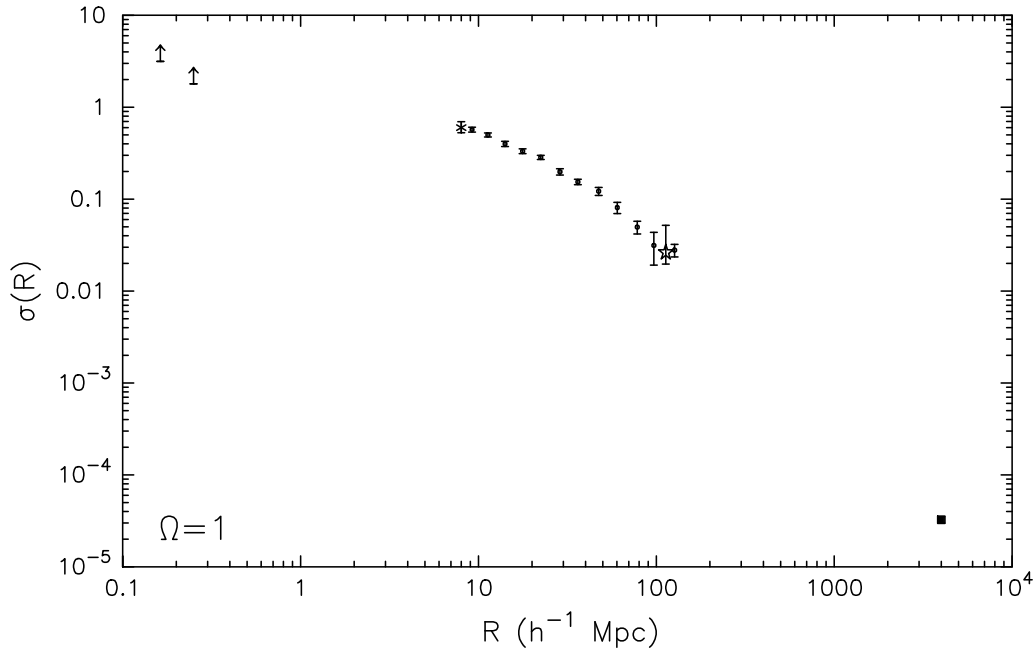}
\caption[figure1]{The observational data, with 1$\sigma$ uncertainties and 
95 per cent confidence lower limits. We represent the {\it COBE} data 
schematically at $4000 h^{-1}$ Mpc; they are indicated by a filled square 
whose size represents the uncertainty. The galaxy correlation 
function data are shown by circles; an uncertainty in overall normalization 
has not been illustrated. The bulk flow constraint is represented by a star, 
and the cluster abundance constraint by a cross. The lower limits are damped 
Lyman alpha systems (left) and quasars (right).}
\end{figure}

While this wide range of parameters gives our analysis a large degree of 
generality, we might as well mention that it is far from exhausting the 
range of possibilities one might contemplate trying. Other approaches, all 
of which would have important consequences, would be to change the number of 
massless particle species, introduce exotic dark matter (degenerate species, 
non-thermal production, warm dark matter, decaying or annihilating dark 
matter), allow exotic inflation giving non-power-law initial perturbations 
or to utilize topological defects (cosmic strings, textures, etc.) as 
the source of initial perturbations. There is no indication from 
observations that any of these are required, but neither can they all be 
excluded.

\section{Observations}

The cited papers give extensive description of the observations we use. 

\begin{figure}[t]
\centering
\leavevmode\epsfysize=9cm \epsfbox{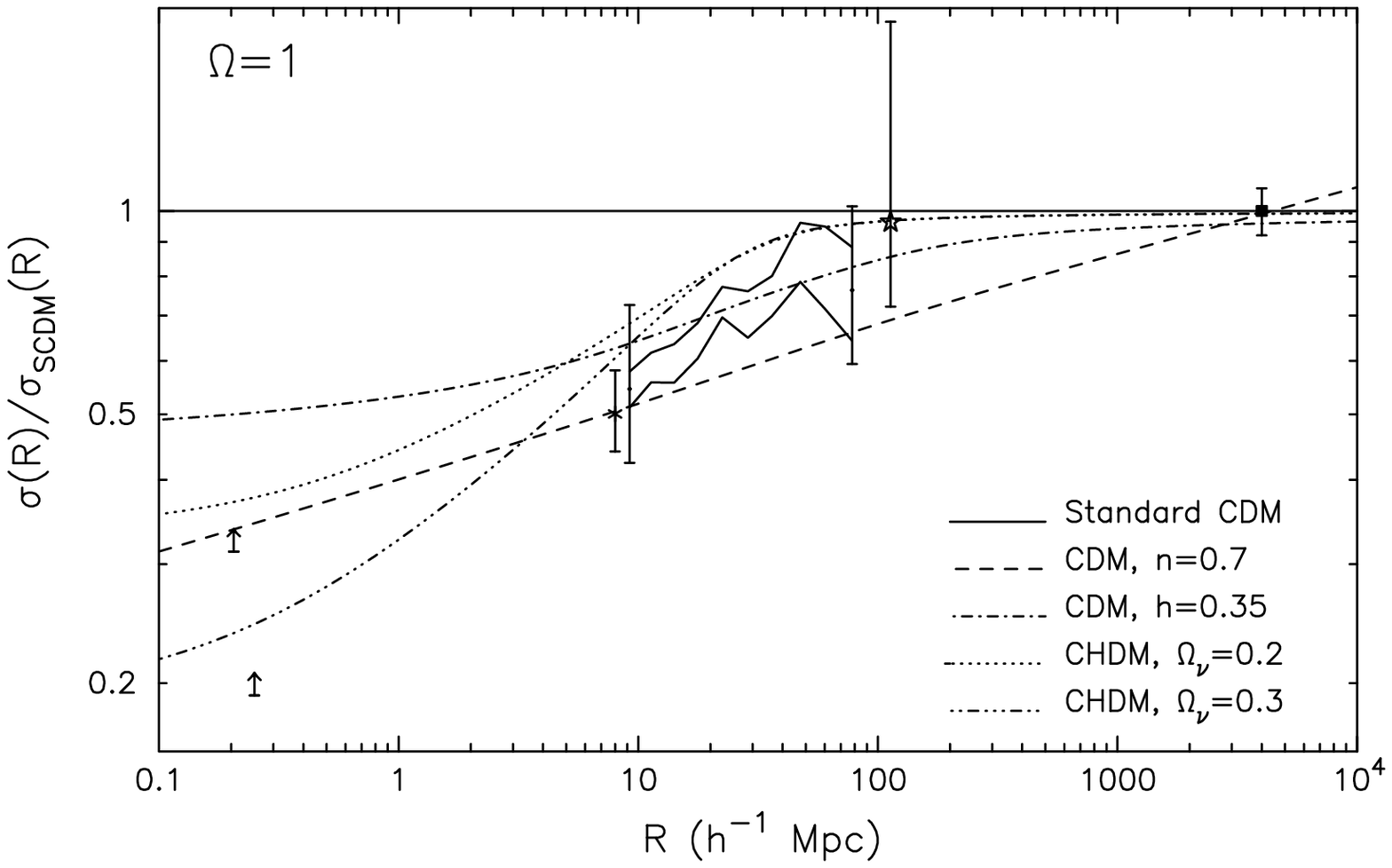}
\caption[figure2]{The data normalized to the prediction of the standard CDM 
model. The solid line is standard CDM; the others modify one 
parameter from this fiducial model, as indicated in the key. All models are 
precisely {\it COBE} normalized; the {\it COBE} point at $4000h^{-1}$ Mpc is 
illustrative. One can shift the entire galaxy correlation function 
data set vertically, corresponding to changing the bias.}
\end{figure}

As stated above, the gravitational waves only affect {\it COBE}. The density 
perturbations, by contrast, are responsible for the complete range of 
structure seen in the Universe. Present observations of microwave 
anisotropies are useful primarily through {\it COBE} on large angular scales 
and via a host of experiments sampling the acoustic (sometimes called 
Doppler) peak at about a degree. Observations constraining the matter power 
spectrum are bulk motions, the shape of the galaxy correlation function and 
the object abundances, the most useful being galaxy clusters today and that 
of quasars, damped Lyman alpha systems and galaxies at high redshift.

To give a feel for some of the data, Fig.~1 shows the observed matter 
power spectrum, under the assumption of a critical density Universe. It is 
shown for illustration only. We plot $\sigma(R)$, the variance of the matter 
distribution smoothed on a scale $R$. It is greatly encouraging to see that 
the data form a smooth curve across orders of magnitude in both 
linear scale and in the size of the perturbations. The normalization of the 
galaxy correlation function depends on the bias parameter; what is crucial 
for us is the shape of this function, which is expected to be 
almost independent of the bias on the large scales used here.

Fig.~2 shows some sample models plotted against this data. Here the 
variance has been normalized by the standard CDM model. Although this model 
clearly doesn't fit the data, it gets us within a factor two or so across 
the relevant scales, allowing us to bring the model into agreement by 
variation of the underlying assumptions. Shown are models with a tilted 
initial spectrum, a very low Hubble parameter or a hot component added. Each 
of these options, corresponding to a single-parameter variation from 
standard CDM, is a reasonable eyeball fit to the data. While it is 
encouraging that the observations are so consistent with one 
another, we see that there is not a huge amount of information contained 
within the matter power spectrum, in comparison to the number of parameters 
we have available to construct models. 

\begin{figure}[t]
\centering
\leavevmode\epsfysize=8cm \epsfbox{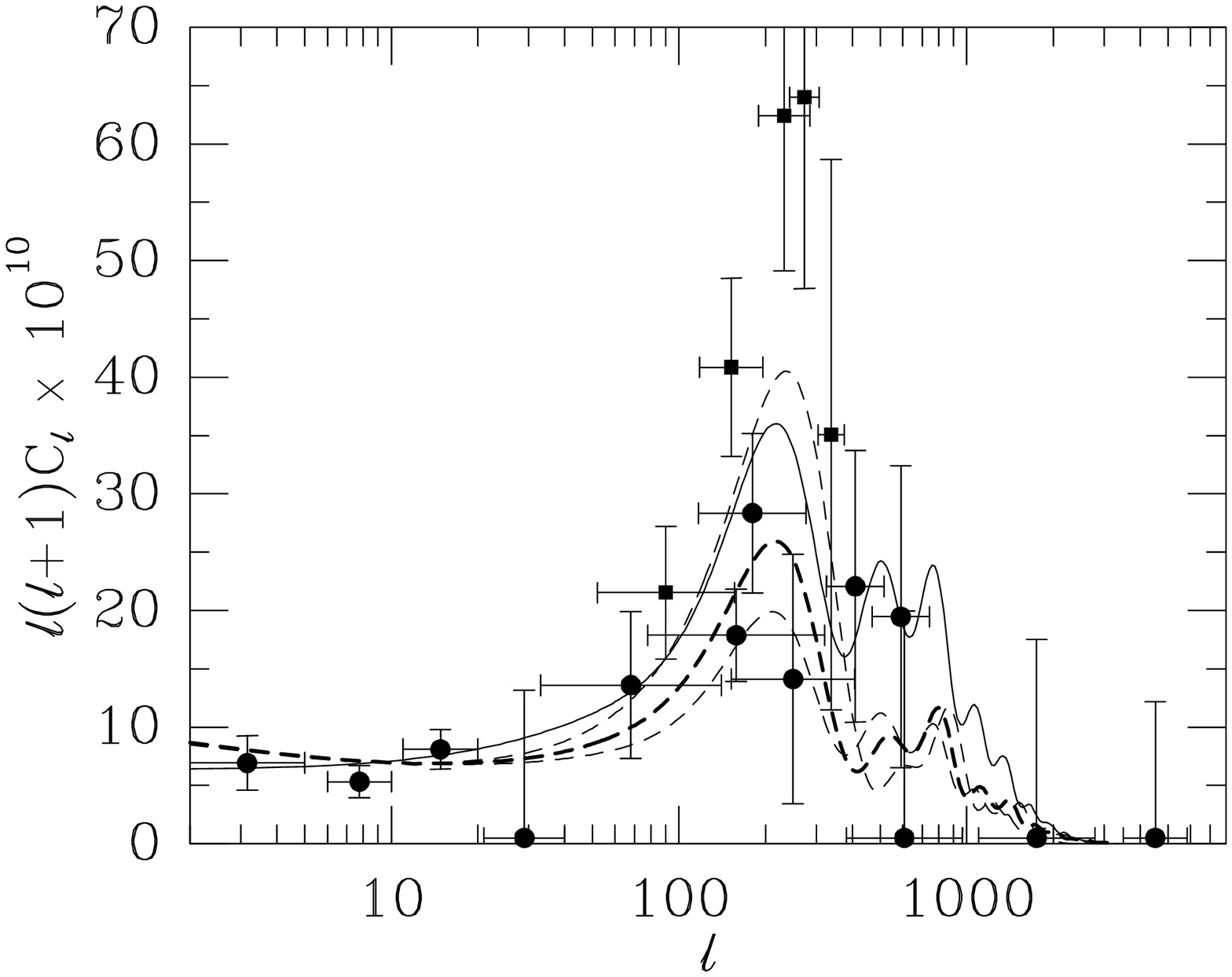}
\caption[figure3]{A view of current microwave anisotropy observations. In 
addition to the large angle anisotropies seen by {\it COBE}, there is 
powerful evidence of a peak on degree scales ($\ell$ of a few hundred). The 
squares are the Saskatoon data, which share a calibration uncertainty. The 
solid line is standard CDM; the remainder are models with $n=0.8$ and 
baryon fractions $\Omega_{{\rm B}} = 0.05$, $0.1$ and $0.2$ (in order of 
increasing peak height).}
\end{figure}

In the future, it seems likely that microwave anisotropy measurements alone 
may give us all the information we need on cosmological parameters. At 
present, the observational situation is beginning to become very interesting 
but one needs to consider both the matter and radiation power spectra to 
draw the strongest conclusions. Fig.~3 shows the present observational 
status. The leftmost four points represent the {\it COBE} four-year 
data \cite{COBE4}, and the rest are from the many experiments probing the 
acoustic peak --- the squares are Saskatoon and the remaining circles, from 
left to right, are SP94, MAX, Python, MSAM, CAT (2 points), White Dish, OVRO 
and ACTA (see Ref.~4 for details).

We see that already there is extremely strong evidence of a peak on degree 
scales. This is a crucial constraint, because it limits the amount of tilt 
permitted in the spectrum; one does not have to tilt very far at all (to 
about $n=0.7$) for the acoustic peak to vanish entirely. This constraint is 
particularly strong on critical-density CDM models, which require a strong 
tilt to give the right matter power spectrum.

\section{Constraints}

\begin{figure}[!t]
\centering
\leavevmode\epsfysize=8cm \epsfbox{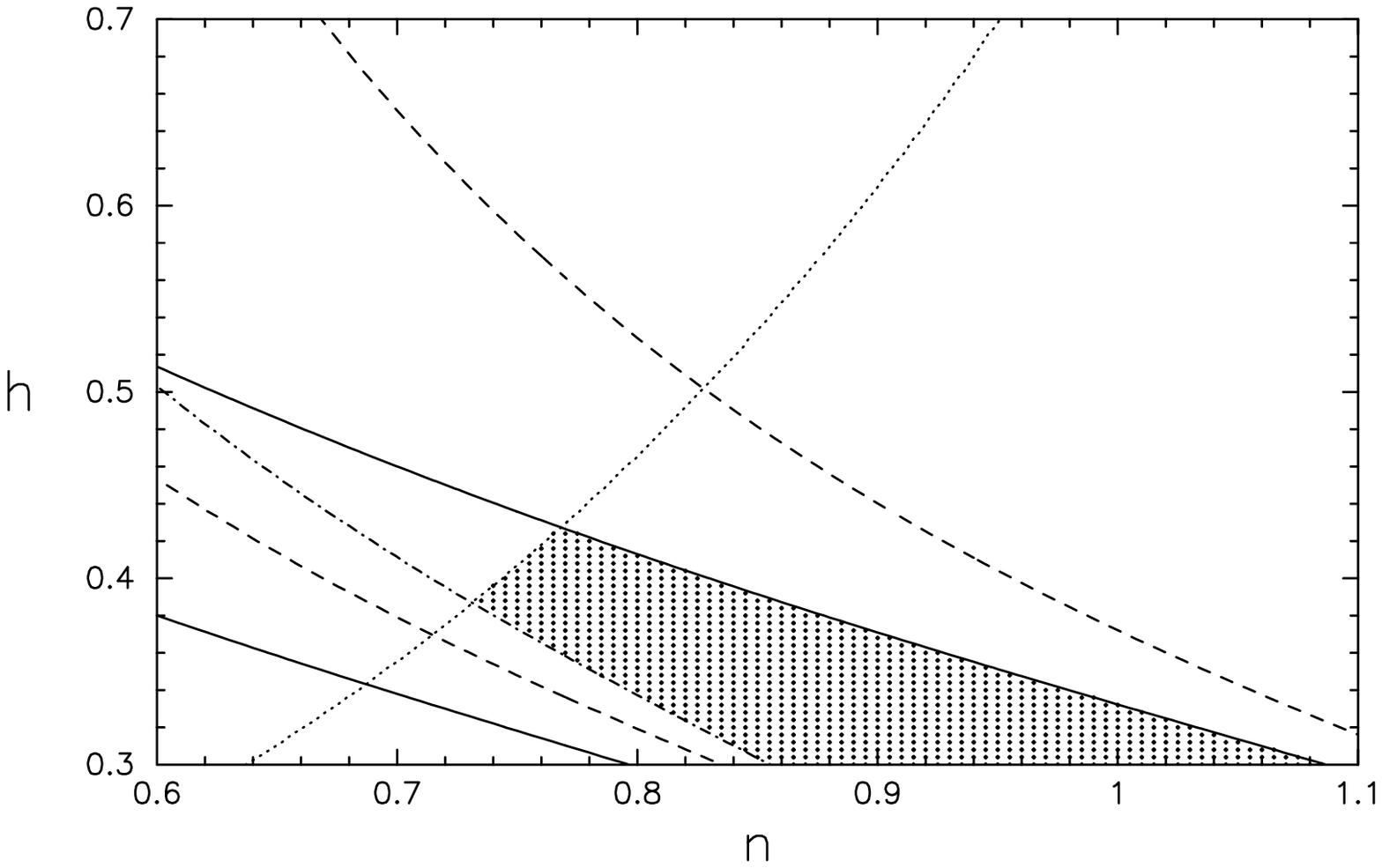}\\
\vspace*{-0.5cm}
\leavevmode\epsfysize=8cm \epsfbox{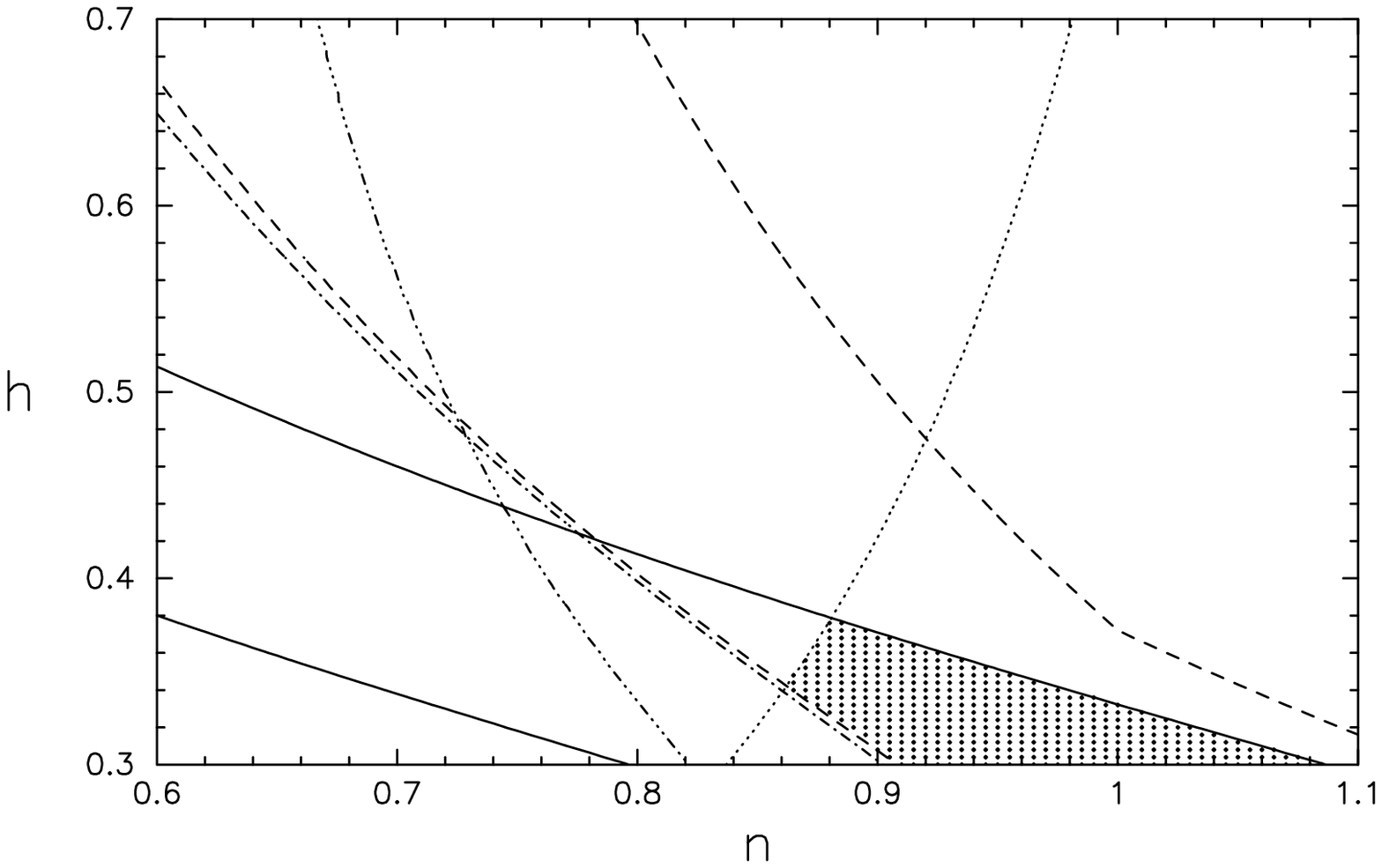}
\caption[figure4]{Critical density cold dark matter models. The top panel is 
without gravitational waves, the lower one includes those as generated by a 
specific model --- power-law inflation. The constraints are: solid, galaxy 
correlation function shape; dashed, cluster abundance; dotted, acoustic 
peak height; dot-dashed, damped Lyman alpha system abundance; 
dot-dot-dot-dashed, galaxy bulk flows. All constraints are plotted at 95\% 
confidence.}
\end{figure}

In this section we give a quick tour of the constraints we find. Much 
greater detail is available in the cited papers.

\subsection{Critical density cold dark matter models}

Fig.~4 shows the $n$--$h$ plane, assuming a baryon density given by 
nucleosynthesis. The strong competition of the acoustic peak and the galaxy 
correlation constraints rules out these models completely, unless one is 
willing to tolerate a very low value of $h$ indeed. Adding gravitational 
waves, as in the lower panel, only makes things worse. 

There seems to be only one escape clause from this predicament (without 
introducing extra forms of matter etc.). That is to raise the baryon density 
above the standard nucleosynthesis range \cite{WVLS}. Baryons have two 
beneficial effects; the acoustic peak is boosted by their pressure (as seen 
in Fig.~3), while they contribute extra damping to reduce power in the 
matter spectrum. Somewhere around 12 or 15 percent baryons seems 
necessary~\cite{WVLS} to permit a fit to the data for $h$ of 0.5, about the 
smallest direct measurement might allow. It is interesting that there would 
independently be a need for such a baryon fraction to explain the observed 
baryon fraction in clusters, and that the range of baryon densities 
permitted by nucleosynthesis extends to higher values in recent analyses.

\subsection{Mixed dark matter models}

\begin{figure}[t]
\centering
\leavevmode\epsfysize=8cm \epsfbox{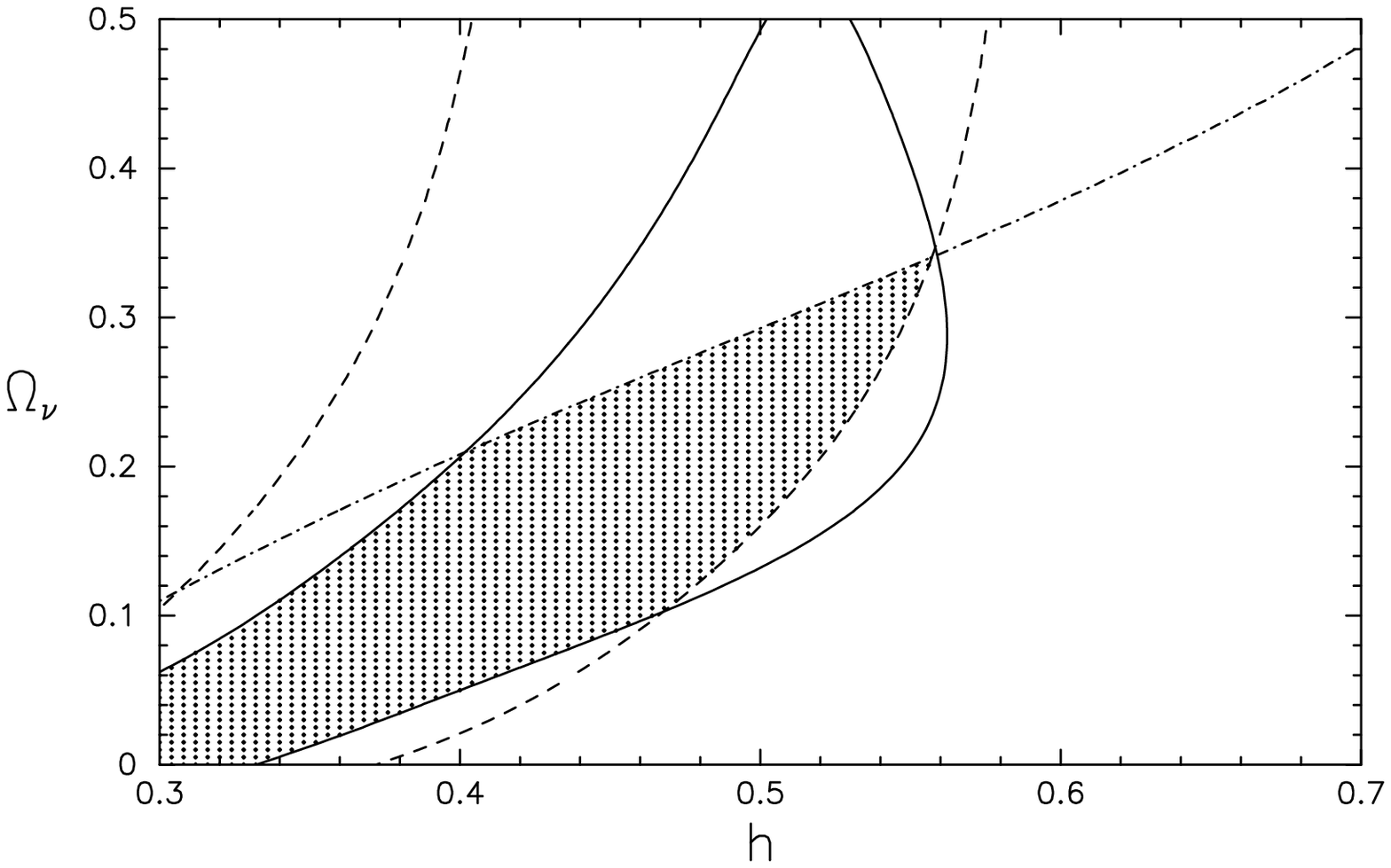}
\caption[figure5]{Mixed dark matter models, without tilt or gravitational 
waves. The constraints are as in Figure 4.}
\end{figure}

Adding a component of hot dark matter is the simplest way to reconcile 
critical-density universes with a higher Hubble constant. Recent 
observational developments aiding these models are the lowering of the {\it 
COBE} normalization between the two and four year data sets, and a weakening 
of the lower limit on the density of gas in damped Lyman alpha systems at 
high redshifts. Even without tilt, as shown in Figure 5, a sizeable allowed 
region exists, all of which produces a satisfactory acoustic peak. 
Introducing tilt allows $h$ up to at least $0.6$.

\subsection{Cold dark matter models with a cosmological constant}

\begin{figure}[!t]
\centering
\leavevmode\epsfysize=8cm \epsfbox{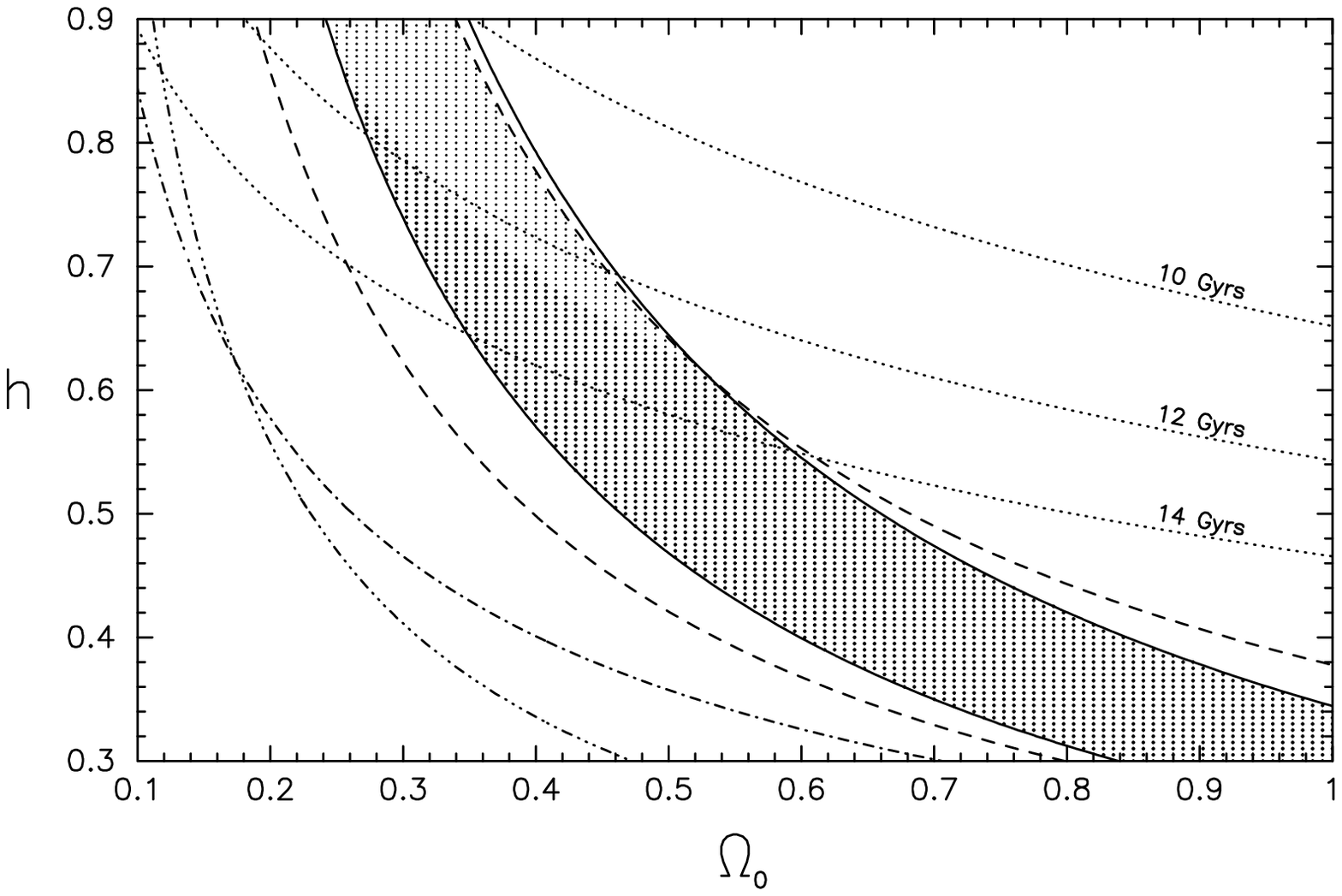}\\
\leavevmode\epsfysize=8cm \epsfbox{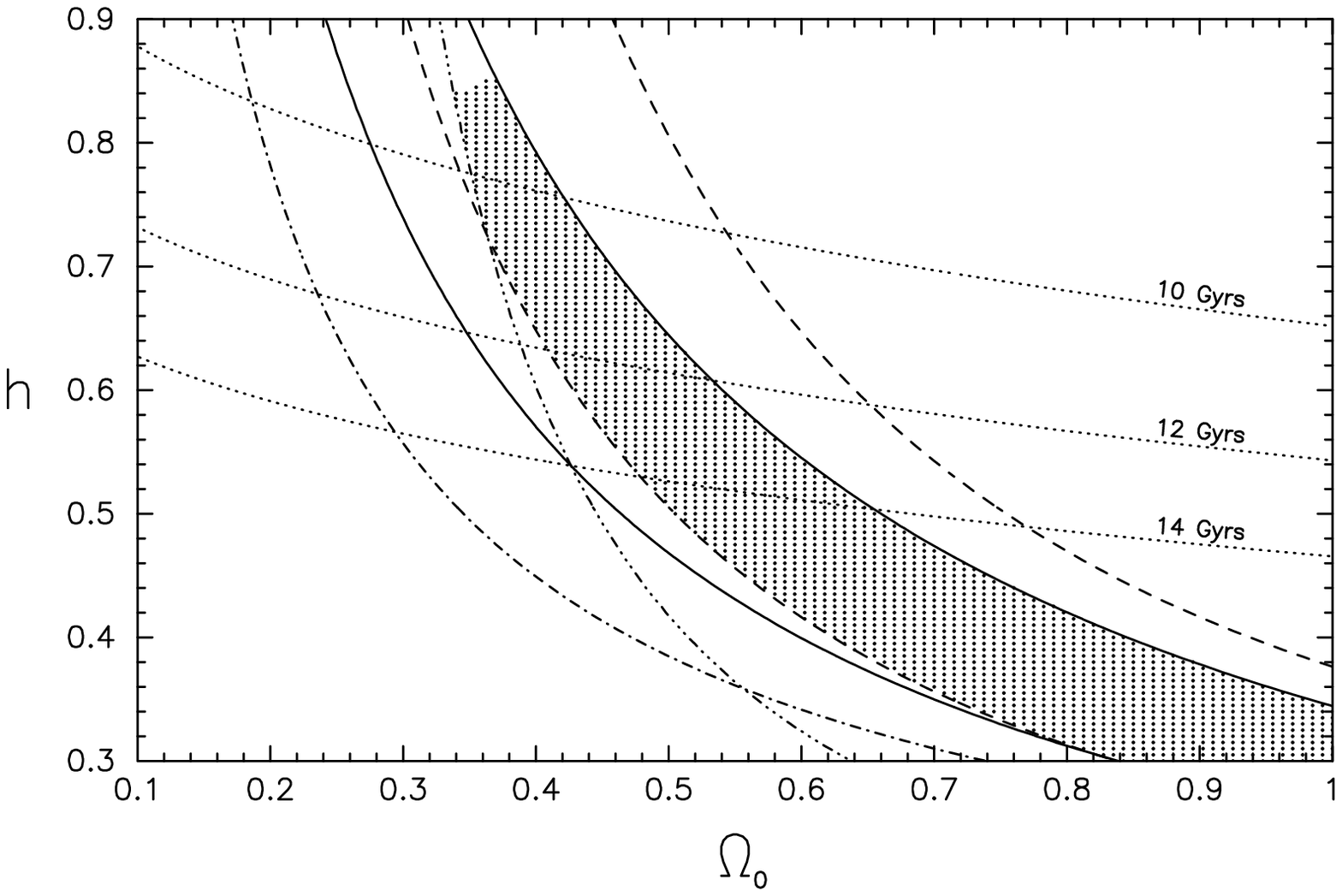}
\caption[figure6]{The top panel is CDM with a cosmological constant, and the 
lower one open CDM. Both cases are without tilt or gravitational waves. The 
dotted lines are of constant age.}
\end{figure}

A truly high $h$, greater than say 0.7, can only be reconciled with the ages 
of objects in a low-density universe. Models with a cosmological constant 
again permit a fit to the data for a wide range of parameters, as in Fig.~6. 
However, one would prefer not to lower $\Omega_0$ much below 0.4; in 
recent years the `fiducial' value of the low density has risen quite a bit. 
At lower densities, the normalization of the power spectrum is such that 
galaxies must be significantly anti-biased, which seems improbable (though 
this conclusion can be weakened by including tilt~\cite{LLVW}). Versions 
including gravitational waves tend to have less viable parameter space.

Note also that, amongst models which fit the data, the low-density models 
are {\em younger}. It is only when one looks at constant $h$ that decreasing 
the density increases the age. An age of 12 Gyrs is probably the lowest the 
community would be willing to live with at this time.

\subsection{Open cold dark matter models}

The (re)discovery of inflation models which give an open universe has 
revived interest in a low-density universe without a cosmological constant. 
Fig.~6 shows the situation without tilt. Again, these models are viable 
provide one keeps $\Omega_0$ above 0.4 or so, with the introduction of tilt 
to $n>1$ lowering this bound.

\section{Conclusions}

\begin{itemize}
\item The use of linear theory is the only way of probing the large 
parameter space to obtain an overall view of what is good and bad. Preferred 
regions can then be subjected to more detailed analysis (see e.g.~Primack 
in these proceedings).
\item All the types of model we've discussed exhibit viable regions:
\begin{description}
\item[CDM:] Much aided by raising $\Omega_{{\rm B}}$. Needs $h \lsim 0.5$.
\item[CHDM:] In good shape, provided $h \lsim 0.6$.
\item[$\Lambda$CDM:] Under pressure from several sources, forcing $\Lambda$ 
down.
\item[Open CDM:] Needs $\Omega_0 \gsim 0.4$.
\end{description}
\item The common ingredient of these inflation-based models is that a 
reasonable amount of cold dark matter always seems to be required.
\item CMB satellites promise to pin down {\em all} these parameters to great 
accuracy. As a by-product, we shall learn much about the mechanism for the 
origin of perturbations.
\end{itemize}

\section*{Acknowledgments}
ARL was supported by the Royal Society and PTPV by the PRAXIS 
XXI programme of JNICT (Portugal). Special thanks to our collaborators David 
Lyth, Dave Roberts, Bob Schaefer, Douglas Scott, Qaisar Shafi and Martin 
White. We acknowledge use of Starlink at the University of Sussex.


\end{document}